\documentclass[sigconf]{acmart}

\usepackage{algorithm}
\usepackage{algorithmic}
\usepackage{booktabs}
\usepackage{multirow}
\usepackage{graphicx}
\usepackage{amsmath}
\usepackage{color}
\usepackage{bbm}
\usepackage{dsfont}

\AtBeginDocument{%
  \providecommand\BibTeX{{%
    \normalfont B\kern-0.5em{\scshape i\kern-0.25em b}\kern-0.8em\TeX}}}

\copyrightyear{2024}
\acmYear{2024}
\setcopyright{acmlicensed}
\acmConference[KDD '24]{Proceedings of the 30th ACM SIGKDD Conference on Knowledge Discovery and Data Mining}{August 25--29, 2024}{Barcelona, Spain}
\acmBooktitle{Proceedings of the 30th ACM SIGKDD Conference on Knowledge Discovery and Data Mining (KDD '24), August 25--29, 2024, Barcelona, Spain}
\acmDOI{10.1145/3637528.3671612}
\acmISBN{979-8-4007-0490-1/24/08}

\begin{document}

\title{ADSNet: Cross-Domain LTV Prediction with an Adaptive Siamese Network in Advertising}







\author{Ruize Wang}
\affiliation{%
  \institution{Tencent Inc.}
  \city{Shanghai}
  \country{China}
}
\email{rayrzwang@tencent.com}

\author{Hui Xu}
\affiliation{%
  \institution{Tencent Inc.}
  \city{Shanghai}
  \country{China}
}
\email{peterhxu@tencent.com}

\author{Ying Cheng}
\authornote{Corresponding authors}
\affiliation{%
  \department{School of Computer Science}
  \institution{Fudan University}
  \city{Shanghai}
  \country{China}
}
\email{chengy18@fudan.edu.cn}

\author{Qi He}
\affiliation{%
  \institution{Tencent Inc.}
  \city{Shanghai}
  \country{China}
}
\email{nickyhe@tencent.com}

\author{Xing Zhou}
\affiliation{%
  \institution{Tencent Inc.}
  \city{Shanghai}
  \country{China}
}
\email{leostarzhou@tencent.com}


\author{Rui Feng}
\affiliation{%
  \department{School of Computer Science}
  \institution{Fudan University}
  \city{Shanghai}
  \country{China}
}
\email{fengrui@fudan.edu.cn}

\author{Wei Xu}
\affiliation{%
  \institution{Tencent Inc.}
  \city{Shanghai}
  \country{China}
}
\email{davidxu@tencent.com}

\author{Lei Huang}
\authornotemark[1]
\affiliation{%
  \institution{Tencent Inc.}
  \city{Shenzhen}
  \country{China}
}
\email{leihuang@tencent.com}

\author{Jie Jiang}
\affiliation{%
  \institution{Tencent Inc.}
  \city{Shenzhen}
  \country{China}
}
\email{zeus@tencent.com}

%

\renewcommand{\shortauthors}{Wang, Xu and Cheng, et al.}

\begin{abstract}
Advertising platforms have evolved in estimating Lifetime Value (LTV) to better align with advertisers' true performance metric which considers cumulative sum of purchases a customer contributes over a period. Accurate LTV estimation is crucial for the precision of the advertising system and the effectiveness of advertisements.
However, the sparsity of real-world LTV data presents a significant challenge to LTV predictive model(i.e., pLTV), severely limiting the their capabilities. 
Therefore, we propose to utilize external data, in addition to the internal data of advertising platform, to expand the size of purchase samples and enhance the LTV prediction model of the advertising platform.
To tackle the issue of data distribution shift between internal and external platforms, we introduce an Adaptive Difference Siamese Network (ADSNet), which employs cross-domain transfer learning to prevent negative transfer. 
Specifically, ADSNet is designed to learn information that is beneficial to the target domain. We introduce a gain evaluation strategy to calculate information gain, aiding the model in learning helpful information for the target domain and providing the ability to reject noisy samples, thus avoiding negative transfer. 
Additionally, we also design a Domain Adaptation Module as a bridge to connect different domains, reduce the distribution distance between them, and enhance the consistency of representation space distribution. 
We conduct extensive offline experiments and online A/B tests on a real advertising platform. Our proposed ADSNet method outperforms other methods, improving GINI by 2$\%$. The ablation study highlights the importance of the gain evaluation strategy in negative gain sample rejection and improving model performance. Additionally, ADSNet significantly improves long-tail prediction. The online A/B tests confirm ADSNet's efficacy, increasing online LTV by 3.47$\%$ and GMV by 3.89$\%$.
\end{abstract}

\ccsdesc[500]{Information systems~Computational advertising}
\ccsdesc[500]{Information systems~Online advertising}

\keywords{Lifetime Value Prediction, Adaptive Cross-Domain Transfer Learning, Computational Advertising}


\maketitle


\begin{figure} [!t]
\centering
  \includegraphics[width=0.98\linewidth]{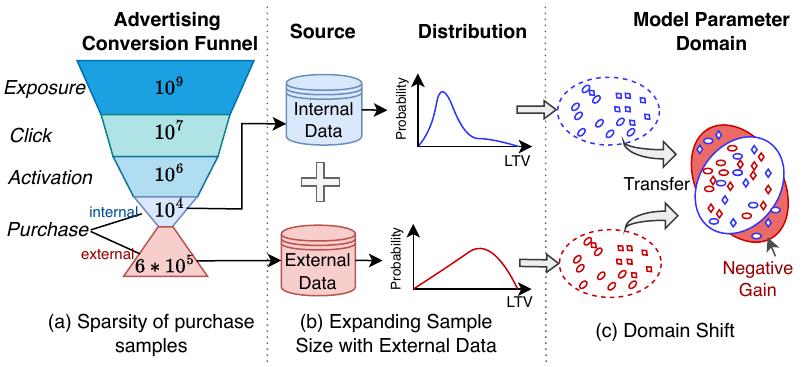}
  \caption{Illustration of the advertising system conversion funnel and challenges, including: (a) the sparsity of internal purchase data, (b) introducing external data, (c) negative transfer due to data distribution shift.
}
  \label{fig:intro}
\end{figure}

\section{Introduction}
The Lifetime Value (LTV) in an advertising system is defined as the cumulative sum of purchases (i.e., revenue for advertisers) a customer contributes over a given period. Given its direct correlation with return on investment (ROI), this metric attracts significant attention from advertisers. Consequently, advertising platforms have progressively evolved to support LTV estimation~\cite{borle2008customer,li2022billion,zhang2023out}, thereby aligning more closely with the assessment requirements of advertisers.
Accurate estimation of LTV plays a crucial role in improving the precision of advertising systems and ensuring the effectiveness of advertisements. 

Recently, some efforts have been made to improve the performance of LTV estimation. For example, \citet{wang2019deep} propose to model LTV as zero-inflated lognormal (ZILN) distribution to address heavy-tailed problem. Some methods~\cite{vanderveld2016engagement,chamberlain2017customer,xing2021learning} propose to model user behaviors and use a feature missing-aware network to reduce the effect of the missing features while training\cite{vanderveld2016engagement,chamberlain2017customer,xing2021learning}. 
However, as purchase is close to the end of the advertising conversion funnel, data sparsity becomes an issue as conversion behavior deepens. This sparsity of real-world LTV data presents a formidable challenge to LTV predictive models, severely limiting their capabilities and having received far less attention.
To address this problem, it is essential to leverage abundant external data to enhance the LTV prediction model of advertising platforms.

Cross-domain transfer learning \cite{pan2009survey,jiang2023forkmerge,zhu2019semi} has emerged as a promising approach to bridge the gap between different domains, particularly when there is a sparsity of labeled data in the target domain. 
This approach has been successfully applied in various fields, such as computer vision, natural language processing, and recommender/adverisiting systems.  
Some methods focus on transferring knowledge from the data perspective through adjustment and transformation of samples and features, including instance weighting\cite{yao2010boosting,jiang2007instance}, feature transformation\cite{xiao2014feature,duan2012learning,li2013learning,wang2017balanced}. Multi-domain learning approaches~\cite{sheng2021one,tang2020progressive} propose to mix multiple sources of data for training a unified model in a multi-task manner. 
However, existing cross-domain transfer learning methods often suffer from negative transfer, which occurs when knowledge of the source domain is not beneficial or even harmful to the learning process of the target domain. Therefore, it is critical to develop a more robust and adaptive cross-domain transfer learning method to mitigate the impact of negative transfer.

In this paper, we propose an \textbf{A}daptive \textbf{D}ifference \textbf{S}iamese Network (\textbf{ADSNet}) to address the challenges associated with LTV estimation and cross-domain transfer learning. ADSNet employs a gain evaluation strategy based on a pseudo-siamese structure, effectively learning beneficial information for the target domain while rejecting noisy samples and avoiding negative transfer. Furthermore, our Domain Adaptation Module bridges different domains, reducing distribution distance and fostering consistent representation space distribution.
We conduct extensive experiments and perform online A/B tests in a real online advertising scenario. Experimental results demonstrate that our approach significantly improves performance on the LTV prediction dataset. Further analysis highlights the effectiveness of our model in rejecting negative gain samples and improving long-tail prediction capabilities.

The contributions of this paper can be summarized as follows:
\begin{itemize}
\item We introduce a novel approach to address data sparsity by integrating external data with the internal data from the advertising system for LTV estimation, utilizing a cross-domain transfer framework.

\item We propose an Adaptive Difference Siamese Network (ADSNet) to tackle negative transfer. Utilizing a pseudo-siamese structure in conjunction with a gain evaluation strategy, we facilitate the assimilation of beneficial external information into the target domain while effectively filtering out noise. Additionally, we incorporate a domain adaptation module to promote consistency across different domains.

\item Extensive experiments reveal that ADSNet surpasses other models, substantially enhancing performance and mitigating negative transfer. Additionally, ADSNet effectively improves long-tail prediction. Online A/B tests further showcase the practical advantages of ADSNet in real-world advertising systems.

\end{itemize}

\begin{figure*} [!ht]
\centering
  \includegraphics[width=0.98\linewidth]{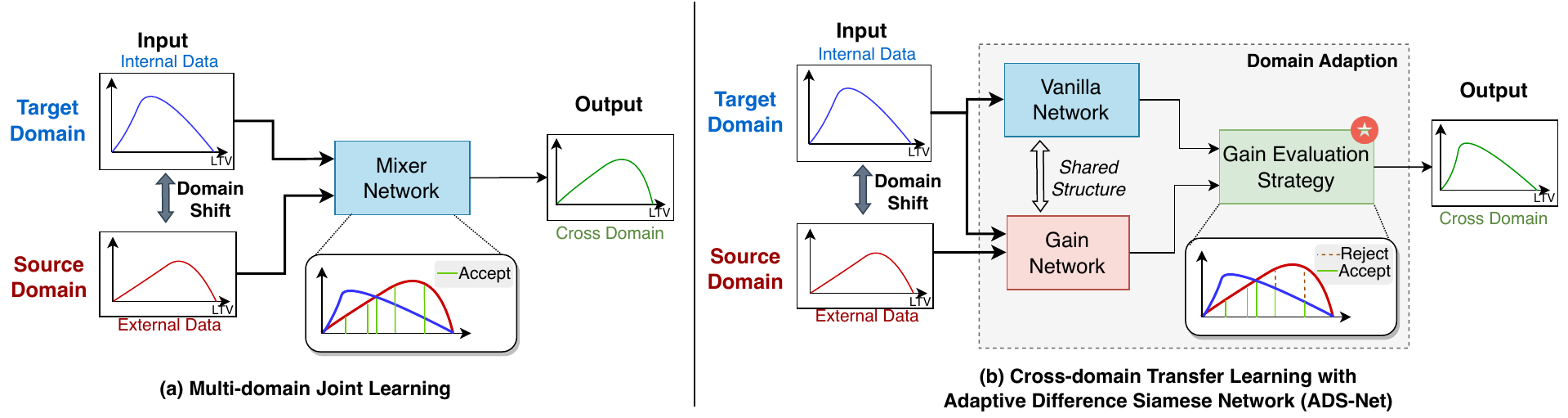}
  \caption{(a) Conventional multi-domain models incorporate source domain knowledge by simply aggregating both source and target domain data and training in a joint manner, which result in introducing noisy samples and even lead to the problem of negative transfer due to data distribution shifts.
  (b) Our ADSNet explores a cross-domain transfer learning method. It employs the pseudo-siamese network to differentially evaluate the information gain provided by the source domain, which supports the rejection of negative gain samples, thereby helping the model learn information that is beneficial to the target domain.
}
  \label{fig:overview}
\end{figure*}

\section{Related Work}
\subsection{Customer Lifetime Value Prediction in Advertising}

Customer Lifetime Value (LTV) prediction has become an integral component of modern advertising platforms, owing to its direct impact on the effectiveness of advertisement placements and overall advertising system precision. Previous studies have focused on various approaches to improve the accuracy of LTV estimation. 
Some previous works~\cite{gupta2006modeling, vanderveld2016engagement, chamberlain2017customer} explore deep learning methods for LTV prediction, emphasizing the potential of neural networks in understanding complex user behaviors. \citet{reddy2022prediction} present a comparative analysis of traditional statistical methods versus machine learning techniques for LTV estimation.
Addressing the issue of missing features in real-world advertising scenarios, MarfNet~\cite{xing2021learning} proposes a feature missing-aware network designed to mitigate the impact of these missing features during training, rather than fill them with default values.
Due to LTV of customers follows a heavy long-tailed distribution with significant fraction of zero value, \citet{wang2019deep} proposes to model the LTV as a zero-inflated lognormal (ZILN) distribution which is described as a mixture of zero-point mass and lognormal distribution, to address the heavy-tailed problem. In this modeling manner, the model fits the mean and deviation of the distribution.
But the distribution assumption is simple and does not meet the multimodal distribution in real scenarios, which leads to limitations. ODMN~\cite{li2022billion} introduces an approach that partitions the complex LTV distribution into several sub-distributions, each trained by distribution experts. 

In the conversion funnel ``\textit{exposure->click->activation->purchase}``, compared to the CTR ``\textit{exposure->click}`` and CVR ``\textit{click->activation}`` prediction, LTV prediction is a more challenging problem due to purchase being the deepest behavior, the data on LTV is very sparse. Although the aforementioned methods obtain better performance on LTV prediction, they struggle with the sparsity of real-world LTV data.
The amount of data determines the upper limit of the model’s performance \cite{wu2023brief,villalobos2022will,vapnik2000nature,bommasani2021opportunities}.
Motivated by such constraints, we utilize external data from sources other than the advertising platform's internal data, which aims to increase the number of purchase samples and improve the LTV prediction model of the advertising platform.

\subsection{Cross-domain Transfer Learning}
Cross-domain transfer learning has emerged as a powerful approach to leverage knowledge from one domain to improve learning in another, especially in scenarios where the target domain suffers from data sparsity. 
However, a significant challenge in this field is the occurrence of negative transfer\cite{pan2009survey,lu2013selective,jiang2023forkmerge}, where irrelevant or noisy information from the source domain hinders the learning process in the target domain, a key challenge in this area.

The concept of domain transfer has also been explored in the context of multi-task learning in Click-Through Rate (CTR) and Conversion Rate (CVR) prediction scenarios\cite{ma2018entire,wang2022escm2}. 
PLE~\cite{tang2020progressive} basically follows the gate structure and attention network for information transfer, similar to MoEs \cite{jacobs1991adaptive,ma2018modeling,zhao2019multiple}. It separates task-sharing and task-specific parameters to learn shared and private representations for each task explicitly, and introduces a progressive routing manner. 
STAR~\cite{sheng2021one} leverages partitioned normalization (PN) to privatize normalization for examples from different domains, and consists of shared centered parameters and domain-specific parameters, adaptively modulating its parameters conditioned on the domain for more refined prediction.
CCTL~\cite{zhang2023collaborative} introduces a framework to mitigate the effects of negative transfer for different business domains in CTR prediction scenario. It evaluates the information gain of the source domain on the target domain using a symmetric companion network.
Compared to CCTL, ADSNet focuses specifically on LTV prediction and employs a gain evaluation strategy to calculate the information gain and reject noisy samples. Additionally, ADSNet introduces a domain adaptation module to reduce the distribution distance between different domains and enhance the consistency of representation space distribution. 

\section{Task Definition}
\textbf{Definition 1 (Lifetime Value).} 
Lifetime Value (LTV) in an advertising system is defined as the cumulative sum of purchases (i.e., revenue for advertisers) a customer contributes over a certain period. This metric is crucial for advertisers in assessing their return on investment (ROI).

\vspace{5pt}
\noindent
\textbf{Definition 2 (Customer LTV Prediction).}
Within the context of an advertising system, the goal of the LTV prediction task is to estimate the LTV $y_i \in Y$ for a given (user, ad) pair $x_i \in X$. In this scenario, $x_i$ represents the specific (user, ad) combination, and $y_i$ corresponds to the LTV generated by the user for the advertisement.
Specifically, given a set of samples $\mathcal{D}=\{x_{i},y_{i}\}_{i=1}^{N}$ consisting of N data-label pairs, we aim to compute the LTV between the user $u_i \in \mathcal{U}$ and an advertisement $a_i \in \mathcal{A}$. This progress can be formulated as following:
\begin{equation}
\hat{y_i}=pLTV_i=f\left(\mathbf{x}_i \mid D, \Theta\right),
\end{equation}
where $\Theta$ denotes the parameters of the LTV estimation model. The model tasks sample $x_i=(u_i,a_i)$ as input, where $u_i$ is user features including user historical behavior (e.g., click and conversion sequences), user profile (e.g., age, gender and purchase frequencey), and $a_i$ is ad features such as the tile and category. $y_i \in[0, \infty)$ is denoted as LTV label. 

\begin{figure*} [!ht]
\centering
  \includegraphics[width=0.9\linewidth]{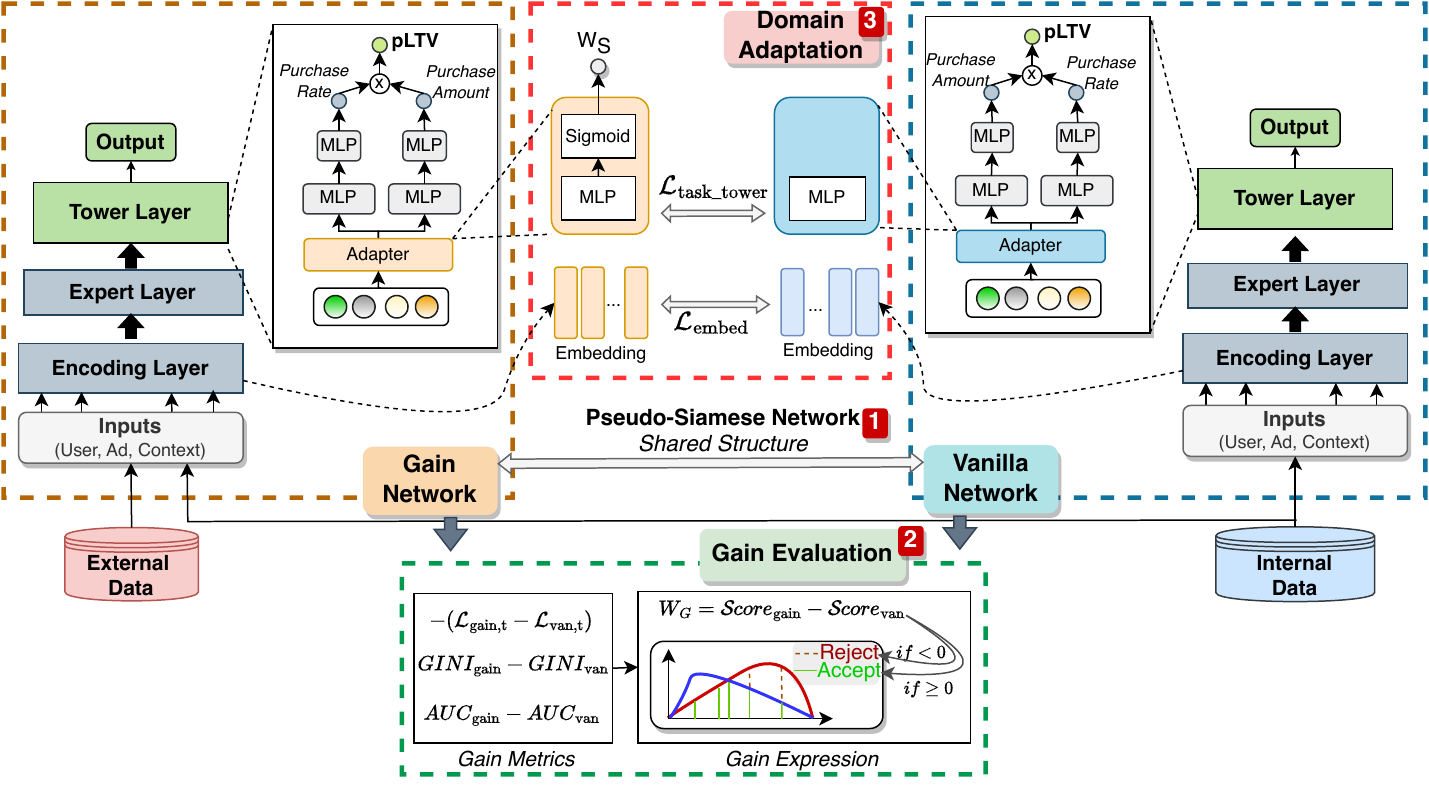}
  \caption{ 
Overview of our proposed ADSNet approach. 
A pseudo-siamese architecture (part 1) is employed to establish a metric to contrast the differences between two networks, allowing the calculation of each input's gain and supporting the rejection of negative gain samples (part 2). A domain adaption module (part 3) is proposed to promote consistency across different domains.
}
  \label{fig:model_structure}
\end{figure*}

\section{Method}
As illustrated in Figure \ref{fig:overview} (a), most of the previous works improve models by integrating knowledge from the source domain via multi-domain joint learning. Regardless of these different variations of methods with multi-domain learning, the common issues not fully studied are negative transfer induced by the domain shift.
To address this, we present our approach ADSNet as shown in Figure \ref{fig:overview}(b) and Figure~\ref{fig:model_structure}. We utilize the pseudo-siamese network to evaluate the information gain, to support learning information that is beneficial to the target domain and reject noisy samples with the gain evaluation strategy. Additionally, we introduce a domain adaptation module as a bridge to connect different domains, reduce the distribution distance between them, and enhance the uniformity of the representation space distribution.
In this section, we first describe the base model(Sec.~\ref{method:base_model}) for LTV prediction, the structure of our ADSNet including the pseudo-siamese network(Sec.~\ref{method:DPSN}), the gain evaluation strategy(Sec.~\ref{method:gain_eval}) and domain adaptation module(Sec.~\ref{method:domain_adap}), and then present the adaptive training process with iterative alignment strategy (Sec.~\ref{method:train_obj}).

\subsection{Backbone for LTV Prediction}
\label{method:base_model}
To better meet the complex distribution of LTV in real advertising scenarios, we develop a deep neural network (DNN) with Ordinal Classification~\cite{cao2020rank,shi2023deep,frank2001simple,fu2018deep,niu2016ordinal} as our backbone for LTV prediction, which is a classic framework consisting of the encoding layer, expert layer, and tower layer. 

\subsubsection{\textbf{Encoding Layer}}
We categorize the features into distinct fields according to their characteristics. For instance, the user's basic profile attributes, such as age, gender, and region, constitute one field, while different user behavior sequences form another field.
Given the input features $x_i$, we encode features in different fields into embedding vectors and employ Field-weighted Factorization Machines (FwFM)~\cite{pan2018field} to model the different interaction of features between different fields, and the embedding vectors are concatenated as the final embedding representation $\mathbf{E}=Encoding(\mathbf{X})$, where $x_i \in \mathbf{X}$ is the input feature.
It is worth noting that alternative encoding methods could also be employed, such as Deep Cross Network (DCN)\cite{wang2017deep}, Learning Hidden Unit Contributions (LHUC)\cite{swietojanski2014learning}, Transformer\cite{vaswani2017attention}, etc., to either replace FwFFM or be used in conjunction with it.

\subsubsection{\textbf{Expert Layer}} 
The expert layer is designed to learn and represent various aspects of the input by incorporating multiple experts, each of which is responsible for capturing specific patterns or characteristics within the data. 
Inspired by the success of Mixture of Experts(MoE) architecture~\cite{jacobs1991adaptive,ma2018modeling}, we employ PLE~\cite{tang2020progressive} as expert layer here, which consists of a set of expert networks and a gating network. Each expert network is implemented as a Multi-Layer Perceptron (MLP), and the gating network is responsible for determining the contribution of each expert to the final output.
Given the encoded representation $e_i \in \mathbf{E}$ from the encoding layer, we feed it into each expert network, obtaining a set of expert outputs $V=\{v_{i}^{j}\}_{j=1}^{K}$, where $K$ is the number of experts. The gating network, which is also an MLP with sofrmax function, takes the same input and produces a set of gating weights $\{g_{j}\}_{j=1}^{K}$, with each corresponding to the weight of expert. The final output $h_{i} \in H$ of the expert layer is a weighted sum of the expert outputs, with the gating weights determining the contribution of each expert:
\begin{equation} 
h_{i} = \sum_{j=1}^K g_j \cdot v_i^j.
\end{equation}

\subsubsection{\textbf{Tower Layer}}
The tower layer takes the expert layer's output and generates the final LTV prediction. The LTV of customers in mobile gaming typically exhibits two distinct traits: 1) a long-tailed distribution with a substantial proportion of zero values, and 2) a multimodal distribution of purchases due to standardized purchase tiers (e.g., \$6, \$30, \$98, and \$198).
To this end, we extend a multi-granularity prediction module, which comprises two components: the probability of purchase prediction at a coarse-grained level and the amount of purchase prediction at a fine-grained level.

\vspace{5pt}
\noindent
\textbf{Probability of Purchase Prediction.}
In particular, we use a classifier to estimate the purchase likelihood of each sample as $p_i$. This classifier is an MLP using a sigmoid activation function. For the prediction of purchase probability, we utilize the cross-entropy loss function. The loss $\mathcal{L}_{prob}$is formally defined as:
\begin{equation}
\mathcal{L}_{prob}=-\mathds{1}\left\{y_i>0\right\} \log (p_i)-(1-\mathds{1}\left\{y_i>0\right\}) \log (1-p_i) ,
\end{equation}
where $\mathds{1}\{\cdot\}$ is the indicator function, which represents the presence of a positive sample, i.e., whether a purchase has been made.

\vspace{5pt}
\noindent
\textbf{Amount of Purchase Prediction.} 
Different from ZILN-based methods~\cite{wang2019deep,yang2023feature} which typically employ a ZILN loss to approximate the complex purchase distribution's mean and variance, we develop a multi-class classification module with ordinal classification~\cite{cao2020rank,frank2001simple,fu2018deep}. It divides the LTV distribution into several sub-distributions and performs prediction over each sub-distribution with multiple binary classifiers. This helps model to learn the ordered nature of purchase categories, and allows for the direct modeling of the cumulative distribution function of purchases, which is more aligned with the inherently sequential progression of purchase amounts. 
We transform continuous purchase labels into a set of binary classification labels to reflect rank information. 
Specifically, the original LTV label $y_i$ is assigned to a segment $s_i$, which represents the segment label for the LTV ranking level. The segments are determined by frequency equalization to maintain a relatively balanced sample size across them. 
Then, each segment(rank) label $s_i$ is expanded to $K-1$ binary class labels $\{s_i^1,\dots,s_i^{K-1}\}$ such that $s_i^k \in \{0,1\}$ is indicates whether $s_i$ exceeds rank $r_k$. For example, $s_i^k=\mathds{1}\left\{y_i>r_k\right\}$. The indicator function $\mathds{1}\{\cdot\}$ is 1 if the inner condition is true and 0 otherwise. 
Each binary classifier employs a sigmoid activation function, and  $p_i^k$ represents the probability prediction of the $k$-th binary classifier.
In the process of inference, the predicted LTV (pLTV) is calculated as:
\begin{equation}
pLTV_i=p_i * \sum_{k=1}^{K-1}\left(p_i^k*\left(\overline{ltv}_k-\overline{ltv}_{k-1}\right)\right),
\end{equation}
where $p_i$ denotes the probability of purchase. $\overline{ltv}_{k}$ is the average LTV of the $k$-th segment. $p_{i}^k$ is the probability of purchase for the $k$-th segment. 

At this stage, we adopt the binary cross-entropy loss for amount of purchase prediction with ordinal classification, and define the purchase amount loss $\mathcal{L}_{amount}$ as follows:
\begin{equation}
\mathcal{L}_{amount} = \mathcal{L}_{\{y_i>0\}}\left[\sum_{k=1}^{K-1}\left(-s_i \log \left(p_i^k\right)-\left(1-s_i\right) \log \left(1-p_i^k\right)\right)\right],
\end{equation}
where $y_i$ denotes original LTV.
To train the above two tasks jointly, we introduce the loss as:
\begin{equation}
\mathcal{L}_{pltv} = \mathcal{L}_{prob} +  \mathcal{L}_{amount}.
\end{equation}

\subsection{Difference Pseudo-Siamese Network}
\label{method:DPSN}
The Siamese network \cite{guo2017learning,dong2018triplet,pulis2021siamese} is a typical architecture in deep learning, which comprises two branches with identical structures and uses similar and dissimilar pairs to learn similarity. In contrast, the Pseudo-Siamese network \cite{gao2020compose,zeng2023multiple} offers more flexibility than the Siamese network, as it allows different structures to receive inputs from various modalities.
Drawing inspiration from these frameworks, we integrate the Pseudo-Siamese Network to assess the information gain, to support learning information from source domain, e.g., external data outside of advertising platform, that is beneficial to the target domain and reject noisy samples. This selective transfer capability is crucial in practical scenarios.

Specifically, our pseudo-siamese network is composed of a vanilla network and a gain Network. Those two networks are based on the backbone for LTV prediction as Sec.~\ref{method:base_model}.
The gain Network receives inputs from both external and internal samples, while the vanilla Network is exclusively fed with samples from the internal channel. 
During training, both networks will update their parameters. This concurrent parameter updating allows each network to learn from its respective data stream, with the gain network adjusting to the nuances of both the external and internal channel samples, and the vanilla network refining its understanding based on the internal channel data alone. This process is key to enabling the pseudo-siamese network to effectively differentiate and integrate relevant information from diverse data sources.
In the training process, we define the losses calculated by $\mathcal{L}_{pltv}$ as $\mathcal{L}_{gain,s}$ of external data and $L_{gain,t}$ of internal data from the gain network. And vanilla network Loss of internal data loss is $\mathcal{L}_{van}=\mathcal{L}_{van,t}$ from vanilla network.

\subsection{Gain Evaluation Strategy}  
\label{method:gain_eval}
Leveraging a pseudo-siamese architecture, we can establish a metric to contrast the differences between two networks, thereby calculating the contribution of input data to the network's performance as:
\begin{equation}
W_G=Score_{gain}-Score_{van},
\end{equation}
where $Score(\cdot)$ is the gain metric function.
In this paper, we utilize this approach to quantify the gain provided by external data to the gain network by examining the variance in losses computed by the two networks on internal samples. Formally, this is expressed as:

\begin{equation}
W_G=\mathcal{L}_{van, t}-\mathcal{L}_{gain, t},
\end{equation}
where $W_G$ represents the gain. If $W_G>0$, it means there is a positive gain to the internal domain.
It is notable that, in addition to the aforementioned methodology, we can extend to employ reinforcement learning by defining metrics that are pertinent to the business objectives. The differential in these metrics can be utilized as a reward signal to train the network by Adversarial Reward Learning~\cite{ho2016generative,wang2018no}.

\begin{algorithm}[t!]
\caption{Iterative Alignment Strategy}
\begin{algorithmic}[1]
\REQUIRE internal samples, external samples, T (total training steps)

\ENSURE $\theta_v$,  $\theta_g$

\STATE \textbf{Warmup Stage:}
\STATE \quad $\theta_v \leftarrow \text{Train}(\text{internal samples})$
\STATE \textbf{Joint Training Stage:}
\STATE \quad $\theta_g \leftarrow \theta_v$
\STATE Set sync\_frequency to 500
\FOR{$t=1$ to $T$}
    \STATE $\theta_g \leftarrow \text{Train}(\text{internal samples} \cup \text{external samples})$
    \STATE $\theta_v \leftarrow \text{Train}(\text{internal samples})$
    \IF{$t \mod \text{sync\_frequency} == 0$}
        \STATE $\theta_v \leftarrow \theta_g$
    \ENDIF
\ENDFOR

\RETURN $\theta_v$,  $\theta_g$

\end{algorithmic}
\label{alg:iterative_alignment_strategy}
\end{algorithm}

\subsection{Domain Adaptation Module}
\label{method:domain_adap}

The domain adaptation module serves as a bridge between the gain Network and the vanilla Network, mitigating the disparity between domain distributions. To this end, an adapter layer is integrated at the bottom of the tower module, implemented as an MLP. 
First, the adapter layer within the gain network is employed to estimate the significance of external data as $W_s$, which is calculated as: 
\begin{equation}
W_{s}=1/exp(sigmoid(MLP(e_i))).
\end{equation}

Then we consider two levels of distributional divergence, encompassing both the low-level embedding layer and the high-level tower layer. Following \cite{jiao2019tinybert}, we employ knowledge distillation to constrain the distributions through Mean Squared Error (MSE):
\begin{equation}
\begin{aligned}
\mathcal{L}_{\text {embed }}&=\operatorname{MSE}\left(\boldsymbol{E}^{V}, \boldsymbol{E}^G\right);\\
\mathcal{L}_{\text {task\_tower }}&=\operatorname{MSE}\left(\boldsymbol{H}_{adapter}^V ,\boldsymbol{H}_{adapter}^G\right),
\end{aligned}
\end{equation}
where $\boldsymbol{E}^{V}$ and $\boldsymbol{E}^{G}$ denote the embeddings from the vanilla network and the gain network, respectively, while $\boldsymbol{H}_{adapter}^{V}$ and $\boldsymbol{H}_{adapter}^{G}$ represent the outputs of the adapter layers within the vanilla network and the gain network, respectively.

The loss of domain adaption is the sum losses as:
\begin{equation}
\mathcal{L}_{\text {domain }}=  \mathcal{L}_{\text {embed }} + \mathcal{L}_{\text {task\_tower }}.
\end{equation}

\subsection{Training Objective}
\label{method:train_obj}
\textbf{Loss Function.} To train the whole model jointly, we introduce a composite loss function $\mathcal{L}_{\text {total }}$ as: 
\begin{equation}
\mathcal{L}_{total}=\mathcal{L}_\text{gain}+\mathcal{L}_\text{van}+\beta*\mathcal{L}_\text{domain}
\end{equation}
where hyperparameter $\beta$ is employed to control the trade-off among these losses. The loss for the gain Nntwork is computed as:
\begin{equation}
\mathcal{L}_{\text{gain}} = W_{s} \cdot \mathcal{L}_{\text{gain, s}} \mathds{1}(W_{G} > 0) + \mathcal{L}_{\text{gain, t}}
\end{equation}
where $\mathcal{L}_{\text{gain, s}}$ is external data loss, $\mathcal{L}_{\text{gain, t}}$ is internal data loss from gain network respectively. $W_s$ is the weight of external data and $W_{G}$ is the information gain. The indicator function $\mathds{1}(W_{G}>0)$ denotes the selective transfer capability of ADSNet, which permits only the advantageous external data to be utilized.

\vspace{5pt}
\noindent
\textbf{Iterative Alignment Strategy.} 
Considering that the performance of the gain network is expected to improve over time, creating a widening gap between it and the vanilla network, such a scenario could eventually render the gain evaluation strategy ineffective. To address this issue, we propose to use iterative alignment strategy during the training process to prevent divergence between the two networks. 
Specifically, the model training is divided into two stages as in Algorithm~\ref{alg:iterative_alignment_strategy}: (1) \textbf{Warmup Stage}: Only internal samples are used to train the vanilla network. This allows the base network to establish a stable starting point. (2) \textbf{Joint Training Stage}: Initialize the gain network with the vanilla network parameters, and train both networks with parameters synchronized iteratively.

\section{Experiments}
\subsection{Experimental Setup}

\subsubsection{\textbf{Datasets}}
Due to the lack of public dataset on LTV prediction, we construct an industry dataset to conduct offline evaluation.
The dataset is collected from from a sampling of conversion logs from Tencent's online advertising system over a span of 90 days, covering four internal traffics (i.e., business domains) and authorized external data from other platforms.
The dataset consists of billions of examples, and is split according to the time axis, with 70, 10, and 10 days' worth of samples allocated for training, validation, and testing, respectively.
The main statistics are shown in Table~\ref{tab:data_stac}, where the sample size is calculated as the average per day.
Due to company privacy policy, we only disclose the purchase sample size and LTV. Table~\ref{tab:data_stac} displays the sample size, and average LTV for each domain. As indicated in this table, different domains exhibit distinct domain-specific data distribution, as reflected in the varying LTVs. It can be observed that the domain with the highest Avg.LTV (Domain \#3) is 4.82, while the domain with the lowest Avg.LTV (Domain \#2) is only 0.07.
The external data have a substantially larger average LTV, indicating a different data distribution from the internal data. These variations in the dataset provide a comprehensive ground for evaluating the performance of LTV prediction models across different domains and data distributions.

\begin{table}[t!]
\caption{Dataset Statistic for LTV Prediction}
\label{tab:data_stac}
\begin{tabular}{@{}ll|cc@{}}
\toprule
\multicolumn{2}{l|}{\textbf{Domain}} & \begin{tabular}[c]{@{}c@{}}\# Sample \\ Size\end{tabular} & Avg. LTV \\ \midrule
\multicolumn{1}{l|}{\multirow{3}{*}{\textbf{Inner Data}}} & Domain1 & 9,136 & 0.34 \\
\multicolumn{1}{l|}{} & Domain2 & 15,706 & 0.53 \\
\multicolumn{1}{l|}{} & Domain3 & 34,129 & 4.82 \\ \midrule
\multicolumn{2}{l|}{\textbf{External Data}} & 225,490 & 8.02 \\ \bottomrule
\end{tabular}
\end{table}

\begin{table*}[ht!]
\caption{Comparison with state-of-the-art LTV prediction approaches.}
\label{tab:compare}
\begin{tabular}{@{}ll|cc|cc|cc|cc@{}}
\toprule
\multicolumn{2}{l|}{\multirow{2}{*}{\textbf{Methods}}} & \multicolumn{2}{c|}{Domain \#1} & \multicolumn{2}{c|}{Domain \#2} & \multicolumn{2}{c|}{Domain \#3} & \multirow{2}{*}{\textbf{\begin{tabular}[c]{@{}c@{}}Average\\ AUC\end{tabular}}} & \multirow{2}{*}{\textbf{\begin{tabular}[c]{@{}c@{}}Average\\ GINI\end{tabular}}} \\ \cmidrule(lr){3-8}
\multicolumn{2}{l|}{} & AUC & GINI & AUC & GINI & AUC & GINI &  &  \\ \midrule
\multicolumn{1}{l|}{\multirow{5}{*}{\textbf{\begin{tabular}[c]{@{}l@{}}Only \\ Internal Data\\ (Single Domain)\end{tabular}}}} & DeepFM~\cite{guo2017deepfm} & 0.719 & 0.731 & 0.692 & 0.633 & 0.9027 & 0.9029 & 0.771 & 0.755 \\
\multicolumn{1}{l|}{} & ZILN~\cite{wang2019deep} & 0.748 & 0.759 & 0.710 & 0.701 & 0.9243 & 0.9237 & 0.794 & 0.794 \\
\multicolumn{1}{l|}{} & FiBiNet~\cite{huang2019fibinet} & 0.753 & 0.762 & 0.721 & 0.712 & 0.9250 & 0.9246 & 0.800 & 0.799 \\
\multicolumn{1}{l|}{} & GateNet~\cite{huang2020gatenet} & 0.749 & 0.767 & 0.743 & 0.732 & 0.9413 & 0.9411 & 0.811 & 0.813 \\
\multicolumn{1}{l|}{} & ADSNet-Backbone (Ours) & \textbf{0.763} & \textbf{0.770} & \textbf{0.757} & \textbf{0.745} & \textbf{0.9472} & \textbf{0.9467} & \textbf{0.822} & \textbf{0.820} \\ \midrule
\multicolumn{1}{l|}{\multirow{4}{*}{\textbf{\begin{tabular}[c]{@{}l@{}}Internal \& \\ External Data\\ (Cross Domain)\end{tabular}}}} & Share-Bottom~\cite{caruana1997multitask} with ZILN & 0.745 & 0.746 & 0.703 & 0.705 & 0.9187 & 0.9201 & 0.789 & 0.790 \\
\multicolumn{1}{l|}{} & MMOE~\cite{ma2018modeling} with ZILN & 0.752 & 0.754 & 0.727 & 0.726 & 0.9249 & 0.9222 & 0.801 & 0.800 \\
\multicolumn{1}{l|}{} & STAR~\cite{sheng2021one} with ZILN & 0.774 & 0.775 & 0.755 & 0.753 & 0.9503 & 0.9448 & 0.826 & 0.824 \\
\multicolumn{1}{l|}{} & CCTL~\cite{zhang2023collaborative} with ZILN & 0.779 & 0.781 & 0.760 & 0.758 & 0.9531 & 0.9522 & 0.831 & 0.830 \\
\multicolumn{1}{l|}{} & CCTL~\cite{zhang2023collaborative} with ADSNet-Backbone & 0.783 & 0.792 & 0.764 & 0.761 & 0.9560 & 0.9537 & 0.834 & 0.836 \\
\multicolumn{1}{l|}{} & ADSNet (Ours) & \textbf{0.792} & \textbf{0.807} & \textbf{0.772} & \textbf{0.771} & \textbf{0.9614} & \textbf{0.9570} & \textbf{0.842} & \textbf{0.845} \\ \bottomrule
\end{tabular}
\end{table*}

\subsubsection{\textbf{Metrics}}
we focus on evaluating the performance of a model that predicts customer lifetime value (LTV) by differentiating high-value customers from low-value ones. For this purpose, we employ two evaluation metrics: the Area Under the Curve (AUC) and the Normalized Gini Coefficient (Norm GINI).
The AUC is a widely used metric for assessing the model's ability to identify purchase users, as it measures the model's classification performance. A higher AUC value indicates that the model can better discriminate between positive (purchase) and negative (non-purchase) classes. However, the AUC does not provide information about the accuracy of the users' ranking based on the predicted LTV. To address this issue, we further adopt the Normalized Gini Coefficient (Norm GINI)~\cite{wang2019deep}. The Norm GINI is a robust measure that captures the model's ability to rank users according to their predicted LTV accurately. 
It is preferred over the Mean Squared Error (MSE) due to its robustness to outliers and better business interpretation.
The Norm GINI ranges between 0 and 1, with a value of 1 indicating perfect consistency between the ranking based on predicted LTV and the ranking based on the real LTV. The lower bound of 0 corresponds to the random ordering of customers.

\subsubsection{\textbf{Hyper-Parameter Settings}}
We implemented all methods using Tensorflow. In the training stage, we utilize the Follow-the-Regularized-Leader (FTRL) optimizer \cite{mcmahan2011follow} for sparse parameters and the Follow the Moving Leader (FTML) optimizer \cite{zheng2017follow} for dense parameters, in which the learning rate is set within the range of \{5e-3, 1e-2\}, respectively.
The batch size is fixed as 512.
The embedding dimension is set to 32, which means that each input feature is represented by a 32-dimensional vector. The architecture of each expert consists of an MLP with two hidden layers with a hidden size of $[$128,64$]$. The tower layer of all methods is designed as an MLP with a hidden size of 32. The weight of the domain adaptation loss, denoted as $\beta$, is set to 0.1 according to the grid search performed on the validation set.

\subsection{Models for Comparison}
We compare our proposed methods with several baselines, including both single-domain and cross-domain settings. Each of these methods is described as follows:

\vspace{5pt}
\noindent
\textbf{Single-Domain.} 
These approaches leverage solely internal data for model training.
\begin{itemize}
\item DeepFM~\cite{guo2017deepfm}: This model integrates the Factorization Machine (FM) model and a deep neural network to extract both low-order and high-order feature interactions. 
\item ZILN~\cite{wang2019deep}: This approach designs a novel zero-inflated lognormal loss to address the imbalanced regression problem. 
\item FiBiNet~\cite{huang2019fibinet}: This model employs a Squeeze-and-Excitation network to discern the significance of feature interactions.
\item GateNet~\cite{huang2020gatenet}: This model utilizes a gating mechanism to distill and select pertinent latent information from feature embeddings.
\item ADSNet-Backbone (Ours): This is the backbone of ADSNet in this paper, characterized by a DNN optimized for ordinal classification.
\end{itemize}

\vspace{5pt}
\noindent
\textbf{Cross-Domain.}  The cross-domain methods integrate both internal data and external data during the training stage, while only validating and testing on internal data. 
\begin{itemize}
\item Share-Bottom~\cite{caruana1997multitask}: This model employs a common architecture which shares the parameters of the bottom layers to facilitate multi-task/multi-domain learning, potentially mitigating overfitting while being sensitive to task discrepancies and data distribution variances.
\item MMOE~\cite{ma2018modeling}: This method builds upon the Share-Bottom approach by adding multiple experts and a gating mechanism to learn the differences between various domains, alleviating the issue of domain variances.
\item STAR~\cite{sheng2021one}: This model consists of shared centered parameters and domain-specific parameters, adaptively modulating its parameters conditioned on the domain.
\item ADSNet (Ours): This denotes our full model, i.e., our backbone with difference pseudo-siamese network and domain adaptation module.
\end{itemize}

\begin{table}[t!]
\caption{Overall metrics on LTV Prediction Task Val Set.}
\label{tab:ablation_study}
\begin{tabular}{@{}ll|l|l@{}}
\toprule
\multicolumn{2}{l|}{\textbf{Method}} & \textbf{AUC} & \textbf{GINI} \\ \midrule
\multicolumn{1}{l|}{1} & ADSNet & \textbf{0.851} & \textbf{0.856} \\ \midrule
\multicolumn{1}{l|}{2} & \textit{w/o Domain Adaptation Module} & 0.833 & 0.835 \\
\multicolumn{1}{l|}{3} & \textit{w/o Gain Evaluation Strategy} & 0.822 & 0.824 \\
\multicolumn{1}{l|}{4} & \textit{w/o Iterative Alignment Strategy} & 0.839 & 0.843 \\
\multicolumn{1}{l|}{5} & \textit{w/o All Components} & 0.819 & 0.820 \\ \bottomrule
\end{tabular}
\end{table}

\subsection{Comparisons with State-of-the-Arts}

Table~\ref{tab:compare} illustrates the performance of various models on the proposed LTV prediction dataset. It is evident that our backbone model (i.e., ADSNet-Backbone) serves as a strong baseline, and our full model (i.e., ADSNet) further enhances performance. 

\begin{figure}[ht!]
\centering
\includegraphics[width=0.48\textwidth]{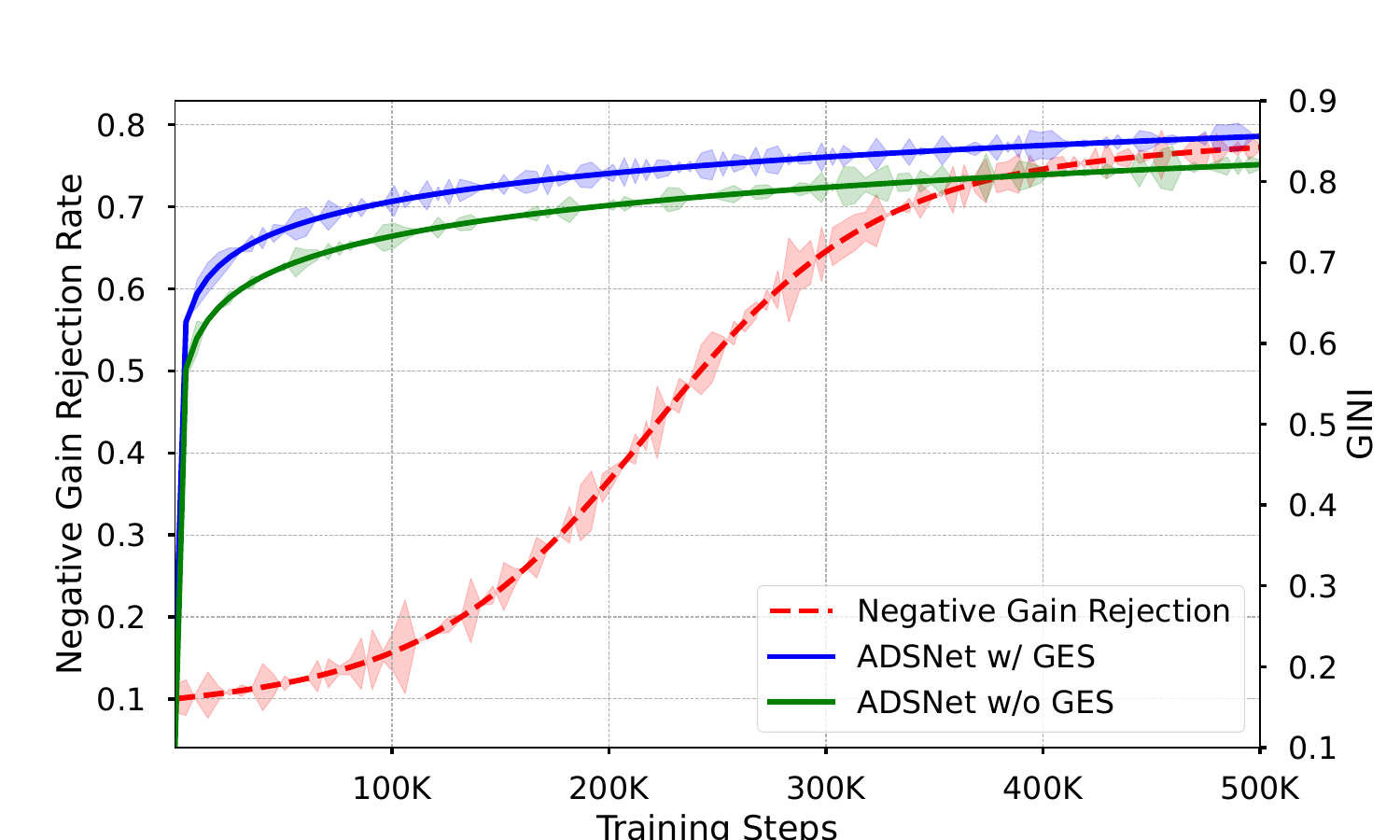}
\caption{Tendency of negative gain rejection rate and GINI during training $w/$or $w/o$ the Gain Evaluation Strategy (GES).}
\label{fig:negative_gain_rejection}
\end{figure}

In particular, our base model outperforms other single-domain methods such as GateNet and ZILN, suggesting that the ordinal classification is more suitable for modeling the complex multi-modal LTV distribution.
Our full model, ADSNet, significantly outperforms cross-domain methods that leverage external data across all domain datasets, achieving absolute overall GINI and AUC improvements of 1$\%$ to 5$\%$. This substantiates the effectiveness of integrating external data into the neural network. 
It is noteworthy that some models experience a decrease in performance when introducing external data. For instance, Share-Bottom w/ZILN, compared to ZILN, shows a decline in overall GINI by 0.4$\%$. This can be attributed to the substantial differences in data distribution across various domains, as illustrated in Table~\ref{tab:data_stac}.
It demonstrates that conventional multi-domain joint learning models fail to avoid the negative transfer phenomenon. In contrast, our ADSNet model exhibits superiority by supporting the rejection of negative gain samples, thus effectively mitigating this issue.

\subsection{Ablation Study}

To investigate the effectiveness and illustrate the impact of different components of our proposed model, we conduct ablation studies on four ablations built upon our full model as shown in Table \ref{tab:ablation_study}. 
In particular, (1) ADSNet~\textit{w/o the Gain Evaluation Strategy} (row 3) exhibits a noticeable decrease in GINI, dropping from 0.856 to 0.824. This suggests that the gain evaluation strategy plays a significant role in rejecting noise samples and stressing negative transfer. (2) The removal of the \textit{Domain Adaptation Module} (row 2) results in GINI dropping by an absolute 2.3$\%$, indicating that constraining the distribution between the gain network and the vanilla network is beneficial to enhance the process of knowledge transfer.
Furthermore, (3) the removal of the \textit{Iterative Alignment Strategy} (row 4) also leads to lower GINI, which indicates that this iterative process is essential for refining the model's predictions in an incremental fashion, aligning the model more closely with the target domain. Iterative alignment likely facilitates a more nuanced adaptation that incrementally bridges the domain gap.

\subsection{Effectiveness of Negative Gain Rejection}
To evaluate the impact of our gain evaluation strategy within ADSNet, we conduct an in-depth analysis to understand the influence of negative transfer during the learning process and to assess the model's ability to identify and reject negative gains. We define the negative gain rejection rate, which represents the probability that external data is rejected, i.e., the gain $W_G<0$.
Figure~\ref{fig:negative_gain_rejection} illustrates the change in the negative gain sample rejection rate and GINI with training steps. The X-axis represents the training steps, the left Y-axis represents the rejection rate of negative gain samples, and the right Y-axis represents GINI. Several observations can be made from this: (1) The GINI exhibits nonlinear growth. It improves rapidly in the early stages and then the rate of improvement slows down over time. (2) The negative gain rejection rate is low in the early stage and then increases with the progress of training. This indicates that the model gradually learns to identify and distinguish negative gain samples during the training process. Ultimately, the negative gain rejection rate stabilizes at 0.64, suggesting that a significant portion of the external samples may not be beneficial if used directly.
These observations highlight the effectiveness of the gain evaluation strategy which ensures that only external data that contribute positively to the target domain are utilized.

\begin{figure}[t!]
\centering
\includegraphics[width=0.48\textwidth]{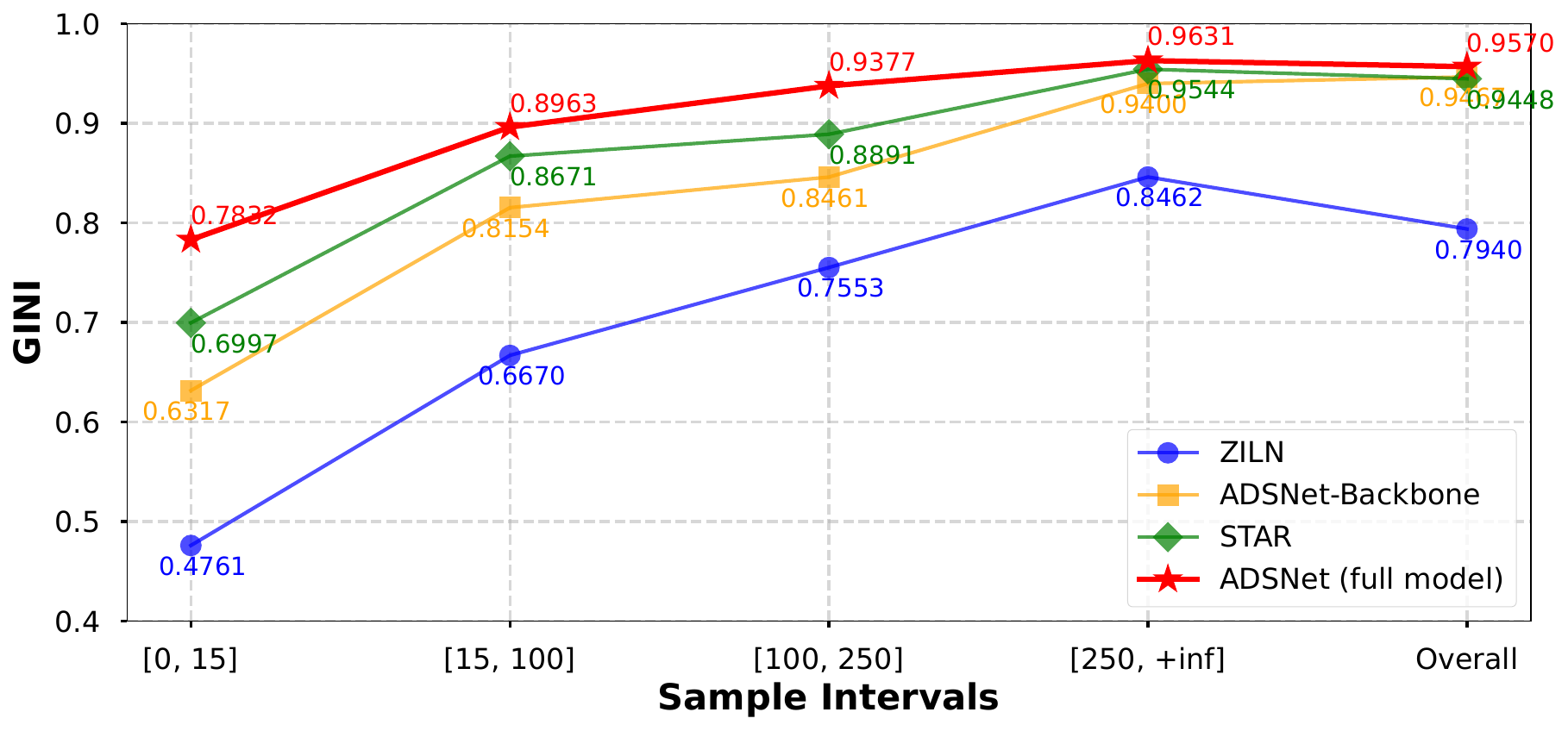}
\caption{Comparison over the interval of sample size.}
\label{fig:long_tail}
\end{figure}

\subsection{Effectiveness of Improving Long-tail Prediction}
In a practical advertising system, we are struggling with the challenge of data sparsity in the LTV estimation scenario. To further understand the improvement achieved by our models, we conduct a more detailed analysis. We quantify the sample sizes corresponding to each advertisement and categorize them into intervals to examine the relationship between the model's predictive capabilities and the size of samples.
The results are presented in Figure~\ref{fig:long_tail}, which demonstrates two insights. First, it intuitively shows that the performance of the model is positively correlated with the sample size. As the sample size increases, the performance of the model improves accordingly. Moreover, compared to the ADSNet-Backbone that does not utilize external data, our ADSNet gains significant improvements by $\sim$15.2$\%$ GINI in the long tail interval $[0,15]$, which indicates that by introducing external data, the model's ability to predict long-tail advertisements is significantly enhanced. 

\subsection{Online A/B Test}
In advertising systems, GMV is a critical outcome metric that cannot be directly optimized. GMV can be represented as GMV = LTV / ROI\_bid. The ROI\_bid is set by the advertiser and is generally fixed based on the required profit margin. Therefore, from a formulaic perspective, GMV and LTV are highly correlated.

We conduct online A/B tests on the real advertising platform of Tencent Ads. Specifically, we employ UV sampling to split the traffic, allocating 5$\%$ of the traffic to the control group and another 5$\%$ to the experimental group. The base serving model is a variant of STAR, adapted to our business characteristics. For the experimental group, we deploy our ADSNet model.
Our online evaluation metrics are the LTV and Gross Merchandise Value (GMV). Due to company privacy policies, we only report the relative improvement. The online A/B test results indicate that ADSNet leads to an increase in online LTV by 3.47\% and GMV by 3.89\% compared with the base model. These results underscore the practical applicability and effectiveness of ADSNet in the LTV prediction task, demonstrating its potential to significantly enhance the performance of real-world advertising systems.

\section{Conclusion and Future Work}
We present the Adaptive Difference Siamese Network (ADSNet) to address the challenges of data sparsity of LTV estimation and cross-domain transfer learning in advertising systems. We incorporate external data to expand the sample size. Our ADSNet utilizes cross-domain transfer learning, a gain evaluation strategy, and a domain adaptation module to learn beneficial information, reject noisy samples, and bridge different domains. Extensive experiments and online A/B tests have demonstrated the effectiveness of ADSNet in improving performance and mitigating negative transfer, as well as its ability to enhance long-tail prediction capabilities.
In future work, we will extend our approach to other related advertising tasks, such as click-through rate prediction to evaluate the generalizability of ADSNet in broader advertising scenarios. 

\begin{acks}
We gratefully acknowledge the contributions of the following: Piao Yang, Ting Wang, Yucheng Hu, Chaoyue Zhao, Liwei Lin, Cong Quan and Kun Bai.
This work was partially supported by the Tencent Rhinoceros Project.
\end{acks}

\bibliographystyle{ACM-Reference-Format}
\balance
\bibliography{sample-base}

\end{document}